\documentclass[aip,jmp,amsmath,amssymb,reprint]{revtex4-1}

\usepackage{color}
\usepackage{enumitem}
\usepackage{array}
\usepackage{float}
\usepackage{graphicx}
\usepackage{dcolumn}
\usepackage{bm}

\makeatletter

\newcommand{\Rmnum}[1]{\expandafter\@slowromancap\romannumeral #1@}
\makeatother

\begin{document}

\title{Alfv\'{e}nic gap eigenmode in a linear plasma with ending magnetic throats}
\author{Lei Chang}
\email{leichang@scu.edu.cn.}
\affiliation{School of Aeronautics and Astronautics, Sichuan University, Chengdu 610065, China}
\date{\today}

\begin{abstract}
To guide the experimental design of a linear plasma device for studying the interaction between energetic ions and Alfv\'{e}nic gap eigenmode (AGE), this work computes AGE referring to fusion conditions in a ultra-long large plasma cylinder ended with strong magnetic throats for axial confinement of charged particles. It is shown that: (i) for uniform equilibrium field between the ending throats, the dispersion relation of computed wave field agrees well with a simple analytical model for shear Alfv\'{e}nic mode; (ii) for periodic equilibrium field with local defect, clear AGE is formed inside spectral gap for both low and high depths of magnetic throats, although lower depth yields easier observation. The strongest AGE can be in order of $3.1\times10^{-4}$ to equilibrium field, making it conveniently measurable in experiment. The AGE is a standing wave localized around the defect which is introduced to break the system's periodicity, and its wavelength is twice the system's period, consistent with Bragg's law. Parameter scan reveals that the AGE remains nearly the same when the number of magnetic ripples is reduced from $18$ to $8$, however, there occurs an upward frequency shift when the depth of magnetic ripples drops from $0.5$ to $0.1$, possibly due to a flute-like effect: shrinking resonant cavity of spectral gap. 
\end{abstract}

\maketitle

\section{Introduction}
\begin{figure}[ht]
\begin{center}$~
\begin{array}{l}
(a)\\
\includegraphics[width=0.4\textwidth,angle=0]{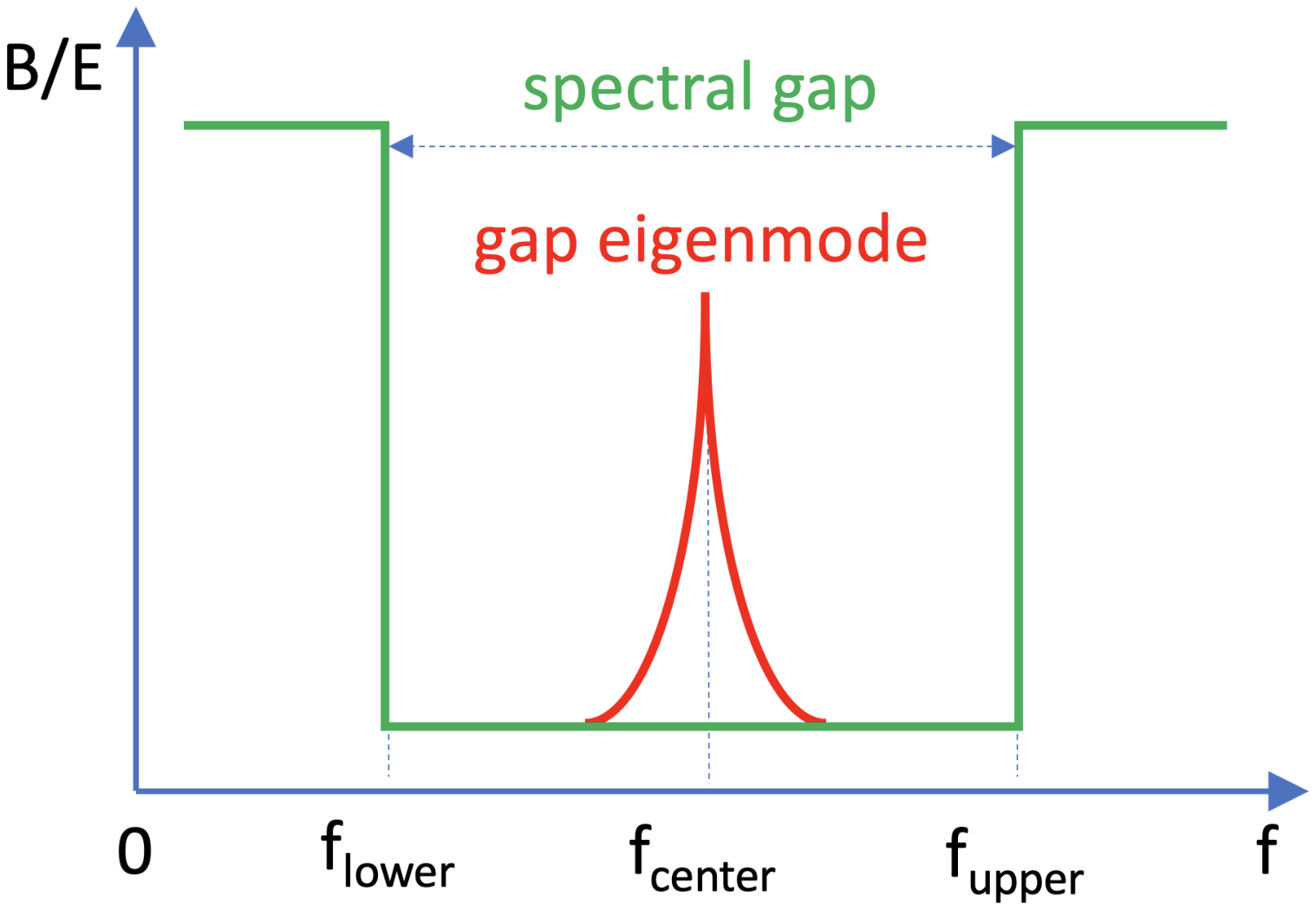}\\
(b)\\
\includegraphics[width=0.4\textwidth,angle=0]{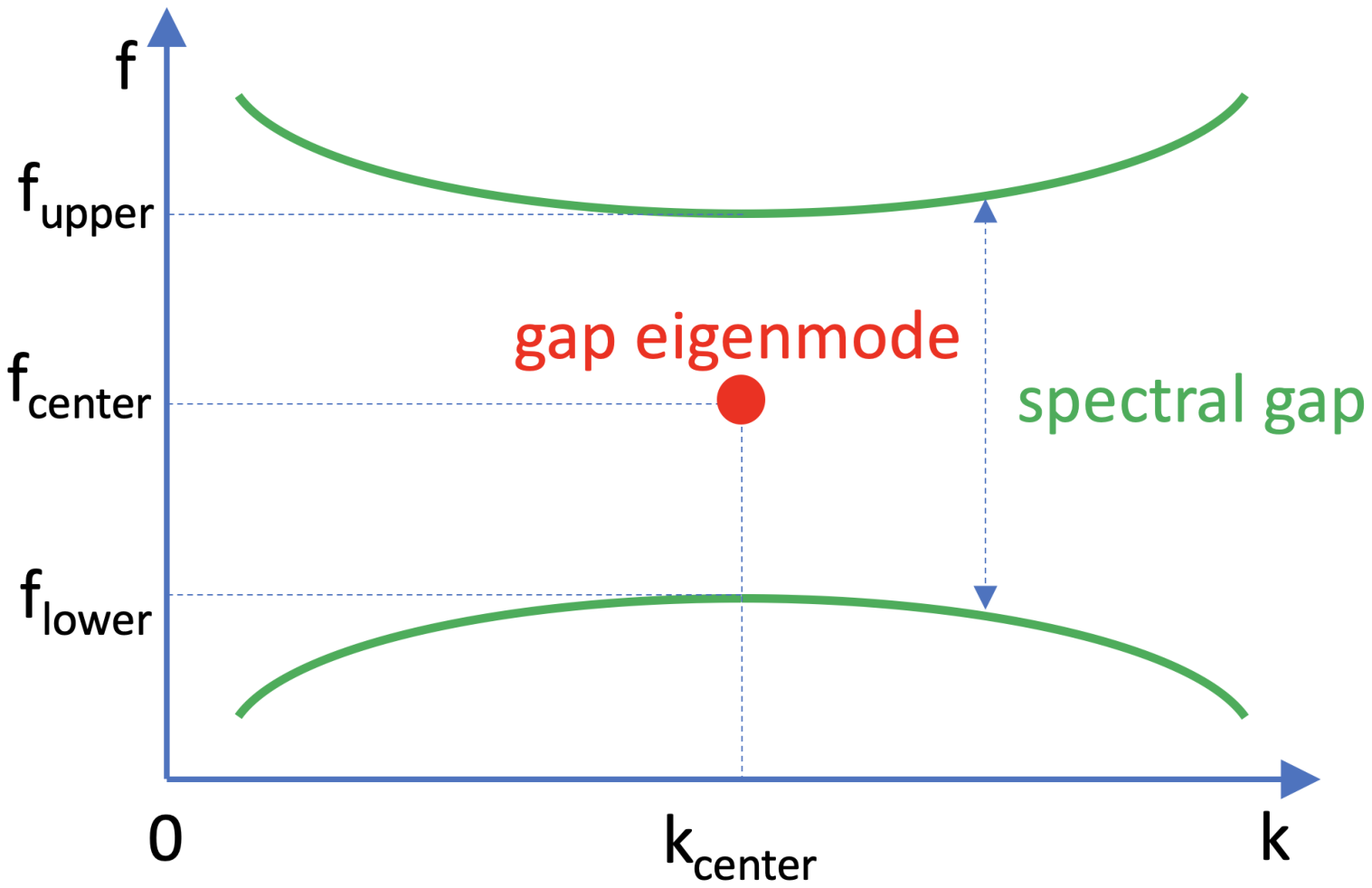}
\end{array}$
\end{center}
\caption{Schematics of spectral gap (green line) and gap eigenmode inside (red line and dot): (a) in the frame of wave field magnitude and frequency, (b) in the frame of frequency and wave propagation number.}
\label{fg10}
\end{figure}

Alfv\'{e}nic instabilities that are driven unstable by energetic particles in fusion plasmas can in turn expel these particles from magnetic confinement, threatening the operation of magnetic fusion reaction.\cite{Duong:1993aa, White:1995aa, Wong:1999aa, Breizman:2011aa} Alfv\'{e}nic gap eigenmode (AGE), a discrete eigenmode inside spectral gap and free of continuum damping, is the most easily excited mode that requires particular attention.\cite{Heidbrink:2008aa, Cheng:1985aa, Fasoli:1996aa} Schematics of general spectral gap and gap eigenmode inside are illustrated in Fig.~\ref{fg10}, which are defined as: a gap in continuous spectrum where waves are evanescent, and a discrete eigenmode with frequency inside the gap, respectively.\cite{Strutt:1887aa, Mott:1968aa, Figotin:1997aa, Kittel:1996aa} Characterizing the interaction physics between AGE and energetic particles especially ions attracts great interest in the past but is mainly based on traditional fusion devices such as tokamak and stellarator.\cite{ITER:1999aa, ITER:2007ab} Because of low cost, simple geometry, easy diagnostic and flexible length, linear plasma is becoming a promising candidate to study fundamental physics occurred in toroidal fusion plasmas, based on its outstanding spatial and temporal diagnostic resolutions.\cite{Gekelman:1991aa, Blackwell:2012aa, Rapp:2017aa}
The attempt to form AGE in a linear plasma can date back to 2008, when Alfv\'{e}nic spectral gap was observed experimentally in a large plasma cylinder, by setting up a longitudinal array of periodic magnetic mirrors according to Bragg's reflection; however, discrete eigenmode was not yet formed inside this gap.\cite{Zhang:2008aa} To guide the experimental implementation of this gap eigenmode, theoretical analyses and numerical computations were carried out, first on whistlers and then on Alfv\'{e}nic modes.\cite{Chang:2013aa, Chang:2014aa, Chang:2016aa, Chang:2016ab, Chang:2018aa} But all these attempts did not consider the loss of energetic particles through the ends of plasma cylinder, which is nevertheless very important for designing a real machine. Recently, ultra-long large plasma cylinder ended with strong magnetic mirrors is proposed by several institutions to study the fundamental physics of the interaction between energetic ions and AGE in fusion plasmas.\cite{Chen:2017aa, Wang:2017aa, Peng:2017aa} The ending mirrors can reflect charged particles many times and eventually confine them, enabling the interaction process between energetic ions and AGE in linear plasma to resemble the scenario in traditional fusion plasmas where energetic ions circle the toroidal geometry all the time.\cite{Freidberg:2007aa, Wesson:2011aa, Artsimovich:1972aa} To assist the design of such a device, the present work constructs deep magnetic throats at the ends of plasma cylinder, with modulation depth much bigger than that of in-between periodic magnetic ripples. It will reveal the minimum number and depth of magnetic ripples required for AGE formation, which is of great importance and practical interest for experimental design, together with the spatial characteristics of AGE. The discovered upward frequency shift for narrowing-down spectral gap may be also helpful to understand the frequency chirping of toroidal Alfv\'{e}n eigenmodes driven by neutral beam ions.\cite{Heidbrink:1995aa, Pinches:2004aa, Zhu:2014aa} Differently, in fusion plasmas energetic ions drive the formation of AGE and then continue to interact with it, whereas in the ongoing construction of linear plasma a local defect is artificially introduced to the system's periodicity to form AGE (present work) and then an energetic ion beam will be shot into the plasma for the latter interaction study. This separated processes could simplify the modeling effort and allow us to focus on the interaction physics in great detail. 

The paper is organized as following: Sec.~\ref{ems} briefly introduces the electromagnetic solver and parameters used to compute the Alfv\'{e}nic wave field, Sec.~\ref{wfd} presents the computed wave field for uniform equilibrium field and AGE in both radial and axial directions, Sec.~\ref{dpd} shows the effect of the number and depth of magnetic ripples on AGE for experimental reference, and Sec.~\ref{smr} gives a short discussion about other potential experimental issues and summarizes the present work. 

\section{Electromagnetic solver}\label{ems}
\begin{figure*}[ht]
\begin{center}
\hspace{0.5cm}\includegraphics[width=0.8\textwidth,angle=0]{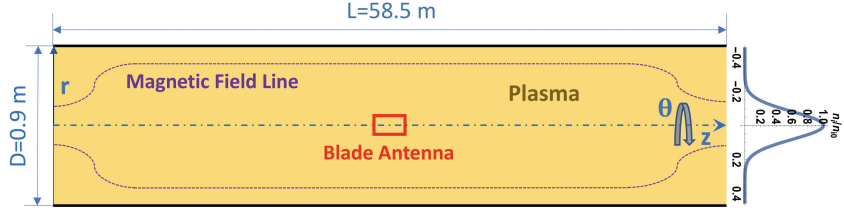}
\end{center}
\caption{A schematic of employed computational domain: rectangle denoting a blade antenna ($L_r=0.005$~m, $L_z=1$~m), dot-dashed line labeling the axis of right-handed coordinate system $(r;~\theta;~z)$, dotted line showing the configuration of uniform field with ending throats, and normalized radial density profile of $n_i(r)/n_{i0}=\exp(-75.1865 r^2)$.}
\label{fg1}
\end{figure*}

To compute the Alfv\'{e}nic waves propagating in an ultra-long large plasma cylinder ended with magnetic throats, the ElectroMagnetic Solver (EMS)\cite{Chen:2006aa} based on Maxwell's equations and a cold plasma dielectric tensor is made use of. The EMS has been employed successfully to interpret experimental data and analytical theory in various studies.\cite{Zhang:2008aa, Chang:2013aa, Lee:2011aa, Chang:2012aa} In the frequency domain, Maxwell's equations are expressed in form of:
\small
\begin{equation}
\nabla\times\mathbf{E}=i \omega\mathbf{B}, 
\end{equation}
\begin{equation}
\frac{1}{\mu_0}\nabla\times\mathbf{B}=-i \omega \mathbf{D}+\mathbf{j_a},
\end{equation}
\normalsize
with $\mathbf{E}$ and $\mathbf{B}$ the wave electric and magnetic fields respectively, $\mathbf{D}$ electric displacement vector ($\partial \mathbf{D}/\partial t=\mathbf{j_{plasma}}+\varepsilon_0\partial \mathbf{E}/\partial t$), $\mathbf{j_a}$ antenna current, $\omega$ antenna driving frequency, and $\mu_0$ the permeability of vacuum. Please note that the other two Maxwell's equations ($\nabla\cdot\mathbf{E}=\rho/\varepsilon_0$ and $\nabla\cdot\mathbf{B}=0$) are independent of frequency but always satisfied by solutions from the EMS. These equations are Fourier transformed with respect to the azimuthal angle $\theta$ and then solved (for an azimuthal mode number $m$) by a finite difference scheme on a 2D domain $(r;~z)$. The quantities $\mathbf{D}$ and $\mathbf{E}$ are linked via a dielectric tensor,\cite{Ginzburg:1964aa}
\small
\begin{equation}
\mathbf{D}=\varepsilon_0[\varepsilon\mathbf{E}+ig(\mathbf{E}\times\mathbf{b})+(\eta-\varepsilon)(\mathbf{E}\cdot\mathbf{b})\mathbf{b}]
\end{equation}
\normalsize
with $\mathbf{b}\equiv\mathbf{B_0}/B_0$ the unit vector along equilibrium magnetic field and 
\small
\begin{equation}
\begin{array}{l}
\vspace{0.15cm}\varepsilon=1-\sum\limits_{\alpha}\frac{\omega+i\nu_\alpha}{\omega}\frac{\omega^2_{p\alpha}}{(\omega+i\nu_\alpha)^2-\omega^2_{c\alpha}},\\
\vspace{0.15cm}g=-\sum\limits_{\alpha}\frac{\omega_{c\alpha}}{\omega}\frac{\omega^2_{p\alpha}}{(\omega+i\nu_\alpha)^2-\omega^2_{c\alpha}},\\
\eta=1-\sum\limits_{\alpha}\frac{\omega^2_{p\alpha}}{\omega(\omega+i\nu_\alpha)}.
\end{array}
\end{equation}
\normalsize
Here different particle species (electron and ion) are labeled by the subscript $\alpha$, and $\omega_{p\alpha}\equiv\sqrt{n_\alpha q_\alpha^2/\varepsilon_0 m_\alpha}$ is the plasma frequency, $\omega_{c\alpha}\equiv q_\alpha B_0/m_\alpha$ is the cyclotron frequency, and $\nu_\alpha$ is a phenomenological collision frequency for each species. Please note that a total effective collision frequency comprising of original Coulomb collision and electron Landau damping is utilized, namely $\nu_{\textrm{eff}}=\nu_{\textrm{ei}}+\nu_{e\textrm{-Landau}}=(n_e e^2/m_e)\eta_{R}+\Lambda v_{\textrm{the}}\omega/v_{A}$, with $\eta_{R}$ the plasma resistivity, $\nu_{\textrm{the}}$ the thermal velocity of electrons and $\Lambda$ an adjusting parameter ($\Lambda=4$ is used throughout this paper according to \cite{Zhang:2008aa}). It is assumed that the equilibrium magnetic field is axisymmetric with $B_{0r}~\ll~B_{0z}$ and $B_{0\theta}=0$, so that near axis expansion can be used for the field, i. e.  $B_{0z}$ is only dependent on $z$,
\small
\begin{equation}
B_{0r}(r,~z)=-\frac{1}{2}r\frac{\partial B_{0z} (z)}{\partial z}. 
\end{equation}
\normalsize
Referring to the experimental work of forming Alfv\'{e}nic spectral gap in a low-temperature plasma cylinder,\cite{Zhang:2008aa} a blade antenna is employed to excite $m=0$ mode. For boundary conditions to determine the wave solution, an ideally conducting chamber is considered so that the tangential components of $\mathbf{E}$ vanish at the walls: 
\small
\begin{equation}
\begin{array}{l}
\vspace{0.2cm}E_\theta(L_r; z)=E_z(L_r; z)=0, \\
\vspace{0.2cm}E_r(r; 0)=E_\theta(r; 0)=0, \\
E_r(r; L_z)=E_\theta(r; L_z)=0,  
\end{array}
\end{equation}
\normalsize
where $L_r$ and $L_z$ are the radius and length of the chamber, respectively; moreover, all field components must be regular on axis. The computational domain is shown in Fig.~\ref{fg1} with employed coordinate system $(r;~\theta;~z)$, uniform equilibrium magnetic field with ending throats, normalized radial density profile $n_i(r)/n_{i0}=\exp(-75.1865 r^2)$, and the central location of blade antenna. The axial density profile is uniform. The radius of the chamber is $0.45$~m, while the length varies from $58.5$~m for uniform field to other values depending on field geometry. Other conditions include ion species of deuterium, electron temperature of $20$~eV, and blade antenna of $L_r=0.005$~m and $L_z=1$~m. 

\section{Computed wave field}\label{wfd}
\subsection{Uniform equilibrium field with ending throats}
\begin{figure}[ht]
\begin{center}
\includegraphics[width=0.45\textwidth,angle=0]{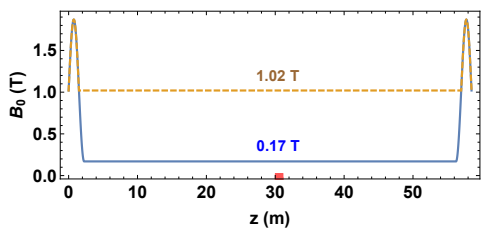}
\end{center}
\caption{Two configurations of uniform equilibrium field with magnetic throats. The field strengths between ending throats is $0.17$~T (solid line) and $1.02$~T (dashed line), respectively. The red bar labels the position of antenna.}
\label{fg2}
\end{figure}

Two configurations of uniform equilibrium field with ending throats are illustrated in Fig.~\ref{fg2}. The ending throats are constructed by $(0.85+0.17)+0.85\sin(2\pi z/3)$, an elevated $\sin$ function with period length of $3$~m. The strengths of in-between equilibrium field are set to be $0.17$~T and $1.02$~T, and the corresponding depths of magnetic throats are $1.87/0.17=11$ and $1.87/1.02=1.83$, respectively. Different depths are chosen to see the effect of ending throat on wave field propagation and thereby gap eigenmode formation. The field strength between ending throats also determines the collisional damping rate and resulted decay length of Alfv\'{e}nic mode, through $3\omega^2\nu_{ei}/(2\omega_{ci}|\omega_{ce}|)$: stronger field leads to longer decay length.\cite{Chang:2014aa, Braginskii:1965aa} To compare the solved wave field in the same frequency range, $n_{i0}=4\times 10^{18}~m^{-3}$ and $n_{i0}=36\times 4\times 10^{18}~m^{-3}$ are chosen for $B_0=0.17$~T and $B_0=1.02$~T respectively, which gives the same phase velocity of Alfv\'{e}nic waves through $v_A=\omega/k_z=B_0/\sqrt{\mu_0 m_i n_i}$. Moreover, the radial density profile of $\exp(-75.1865 r^2)$ is configured to shrink most plasma within $r\leq 0.3$~m for weakening the interference of radial wall. 

\begin{figure*}[ht]
\begin{center}$
\begin{array}{ll}
(a)&(b)\\
\includegraphics[width=0.45\textwidth,angle=0]{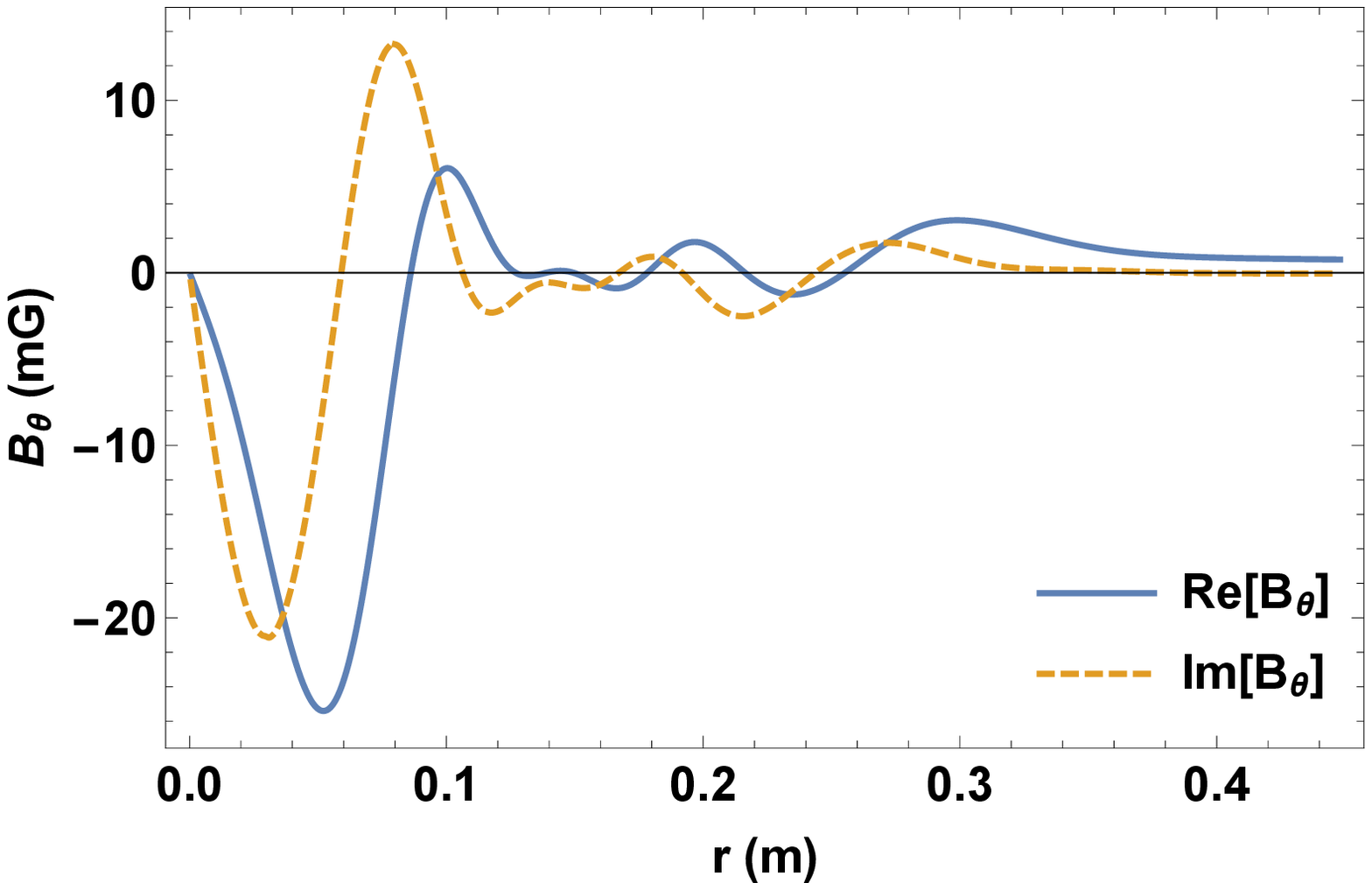}&\includegraphics[width=0.45\textwidth,angle=0]{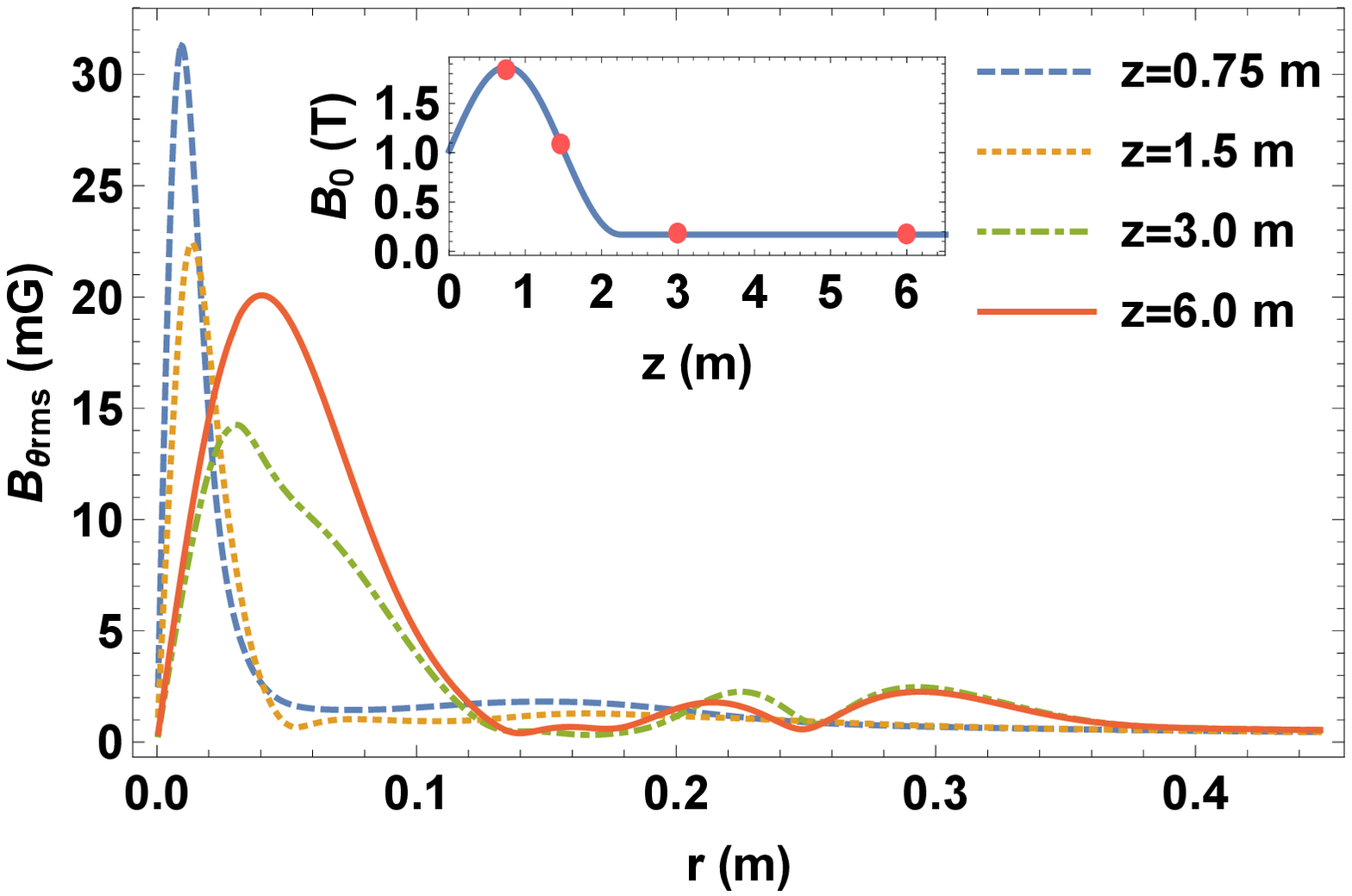}\\
(c)&(d)\\
\hspace{0.2cm}\includegraphics[width=0.45\textwidth,angle=0]{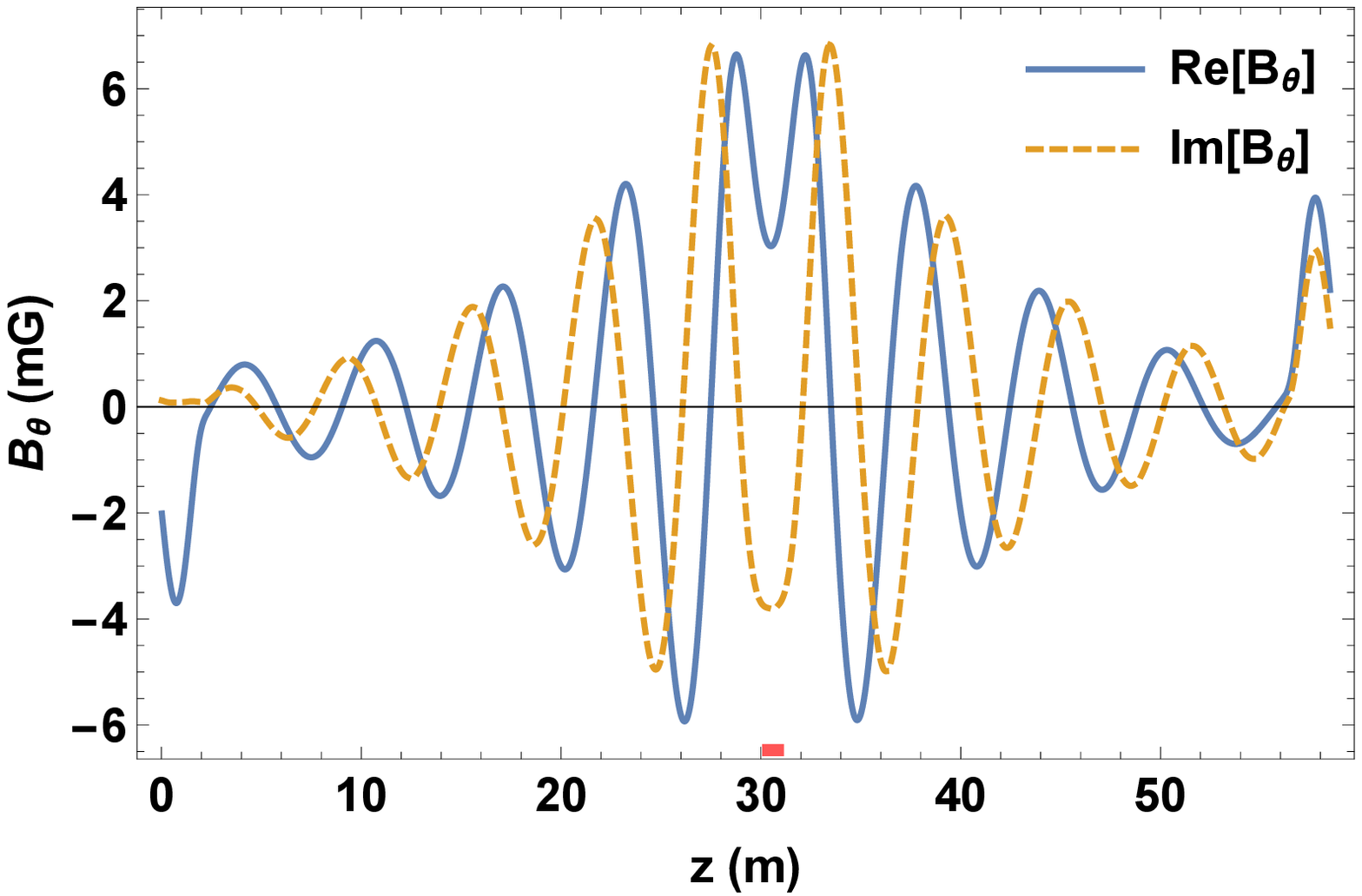}&\includegraphics[width=0.45\textwidth,angle=0]{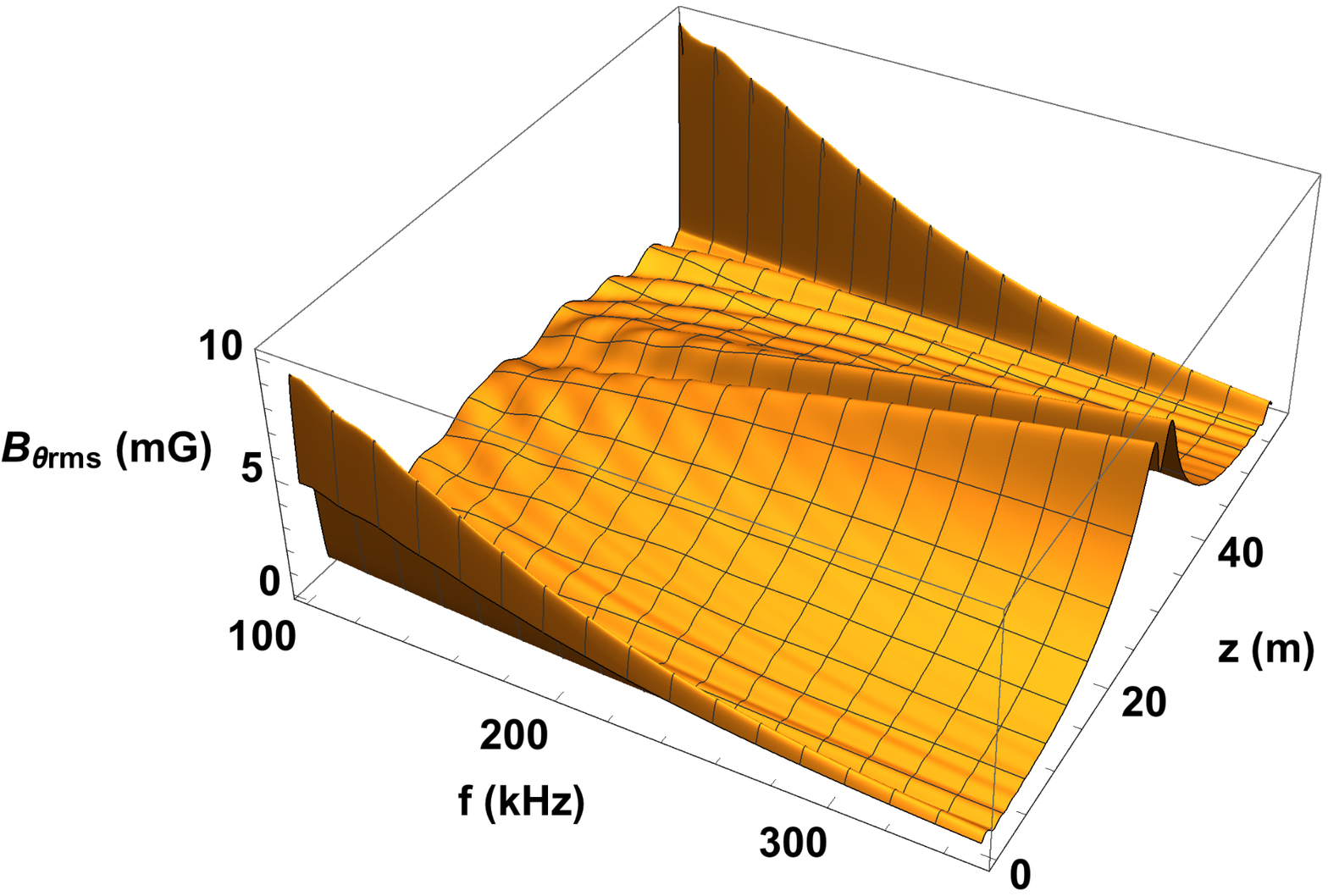}
\end{array}$
\end{center}
\caption{Variations of wave magnetic field for $B_0=0.17$~T: (a) real and imaginary components of $B_\theta$ for $f=225$~kHz at $z=6$~m, (b) radial profiles of $B_{\theta rms}$ for $f=225$~kHz at different axial locations (marked by red dots in the inset figure), (c) real and imaginary components of $B_\theta$ for $f=225$~kHz at $r=0$~m, (d) surface plot of $B_{\theta rms}$ in the space of $(f; z)$ at $r=0$~m. The red bar labels the position of antenna. }
\label{fg3}
\end{figure*}

\begin{figure*}[ht]
\begin{center}$
\begin{array}{ll}
(a)&(b)\\
\includegraphics[width=0.44\textwidth,angle=0]{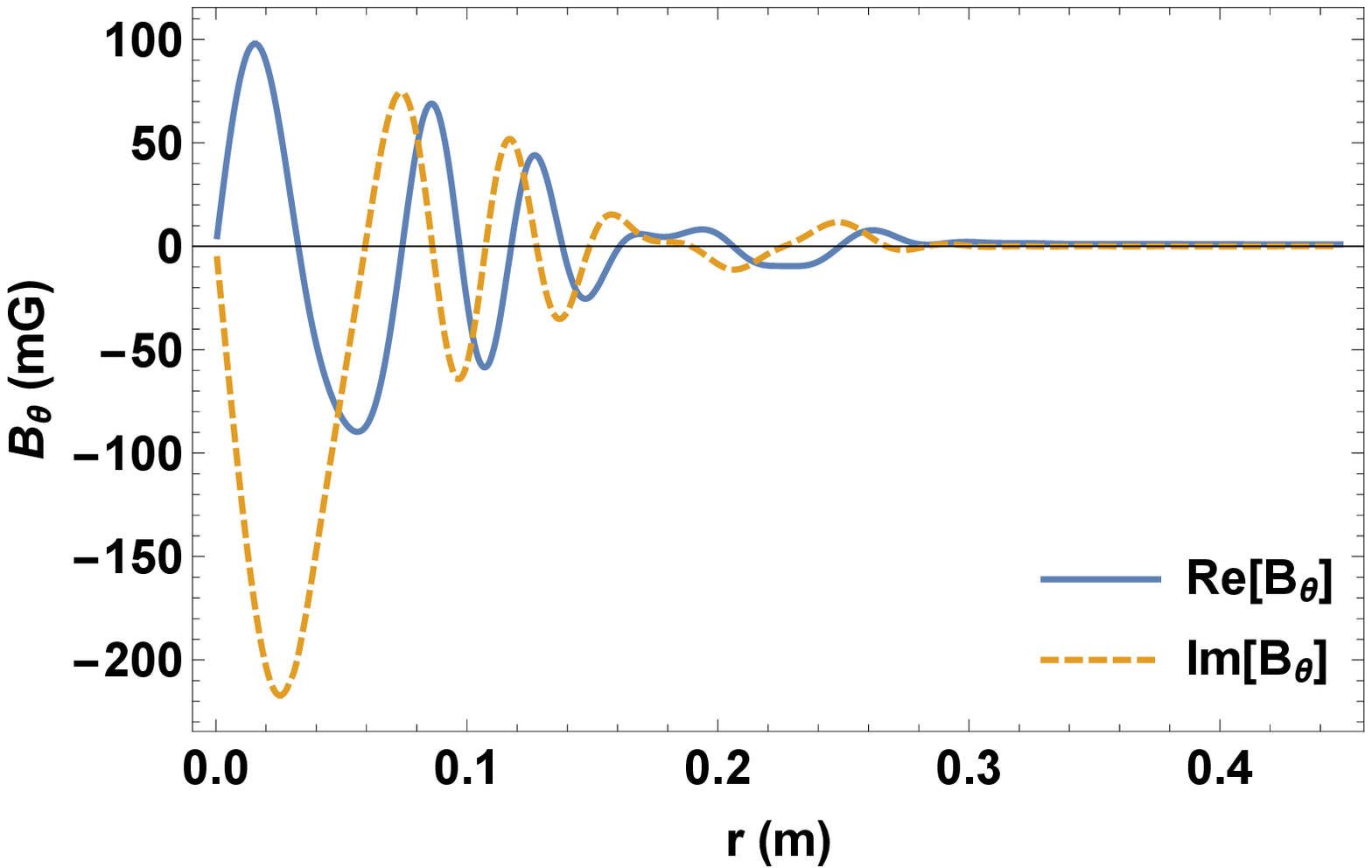}&\includegraphics[width=0.45\textwidth,angle=0]{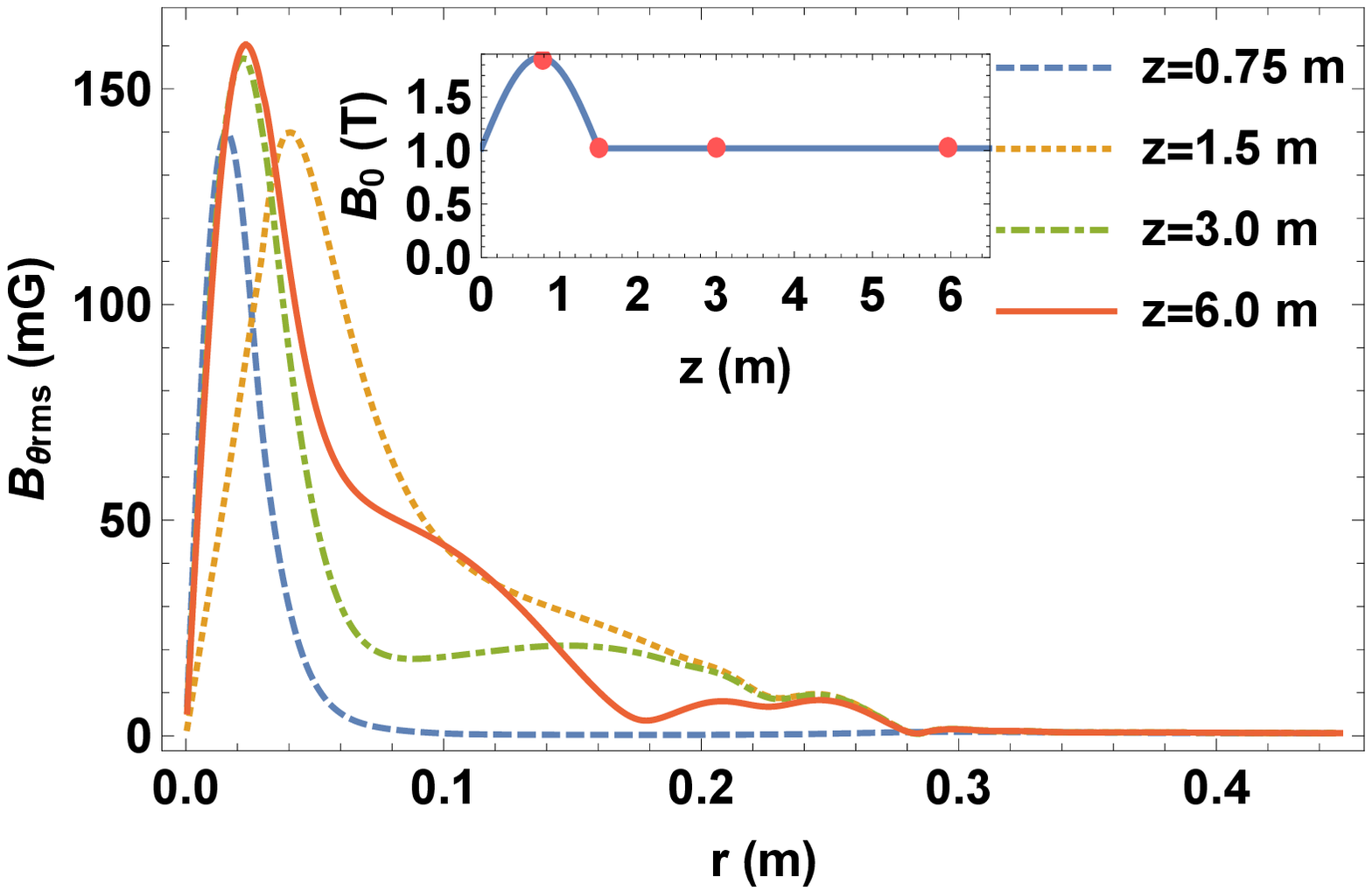}\\
(c)&(d)\\
\hspace{0.2cm}\includegraphics[width=0.45\textwidth,angle=0]{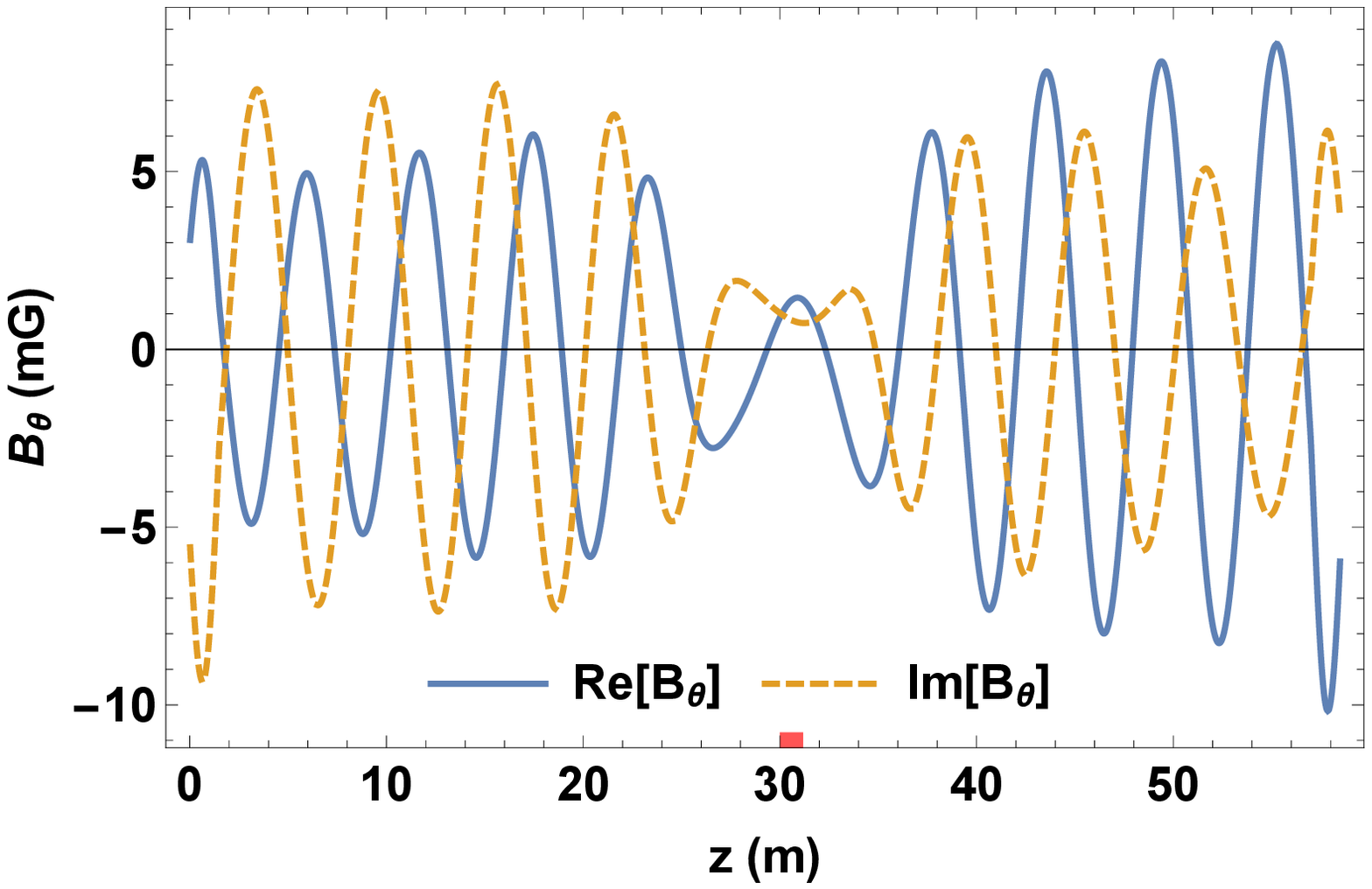}&\includegraphics[width=0.45\textwidth,angle=0]{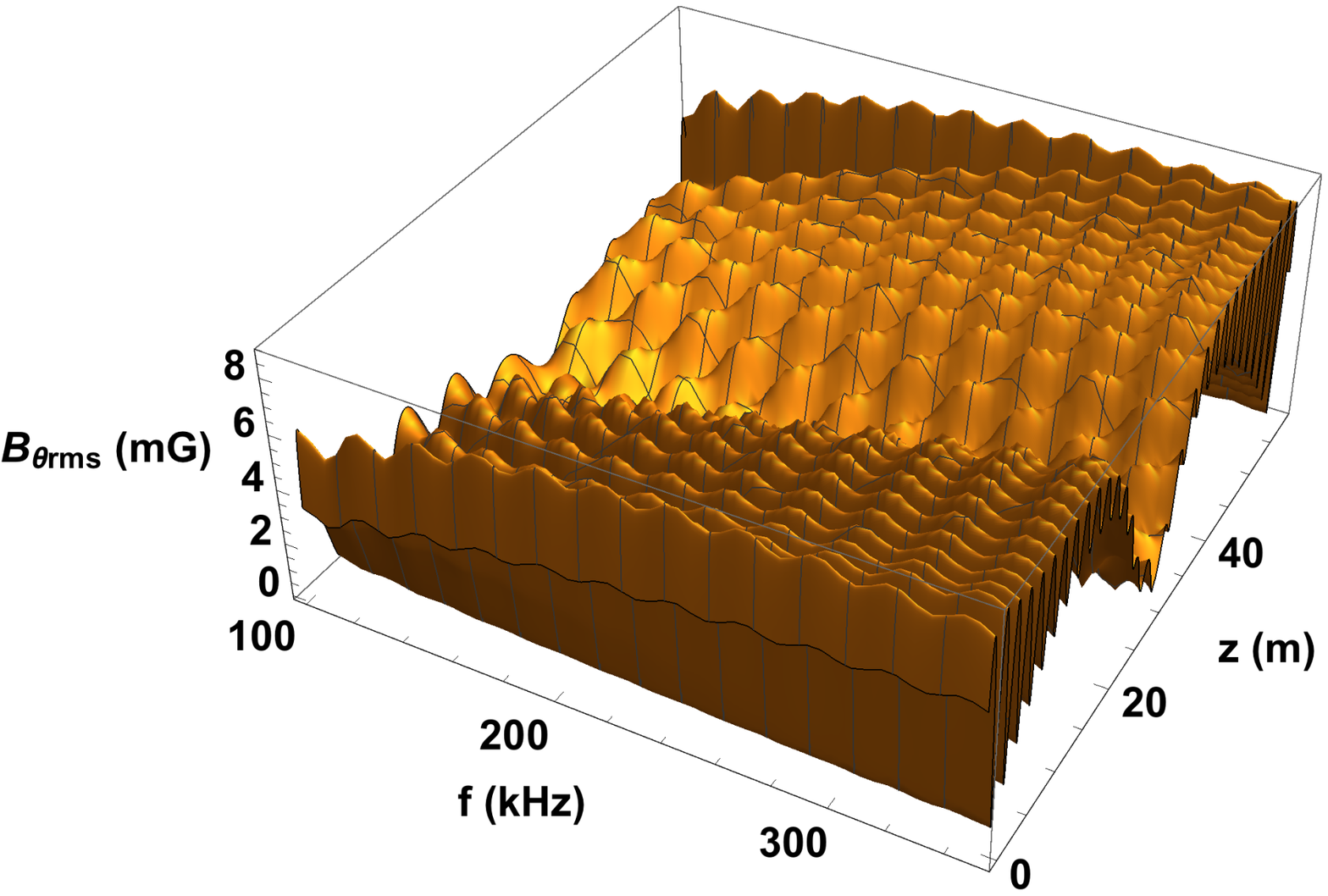}
\end{array}$
\end{center}
\caption{Variations of wave magnetic field for $B_0=1.02$~T: (a) real and imaginary components of $B_\theta$ for $f=225$~kHz at $z=6$~m, (b) radial profiles of $B_{\theta rms}$ for $f=225$~kHz at different axial locations (marked by red dots in the inset figure), (c) real and imaginary components of $B_\theta$ for $f=225$~kHz at $r=0$~m, (d) surface plot of $B_{\theta rms}$ in the space of $(f; z)$ at $r=0$~m. The red bar labels the position of antenna.}
\label{fg4}
\end{figure*}

\begin{figure}[ht]
\begin{center}$
\begin{array}{l}
(a)\\
\includegraphics[width=0.45\textwidth,angle=0]{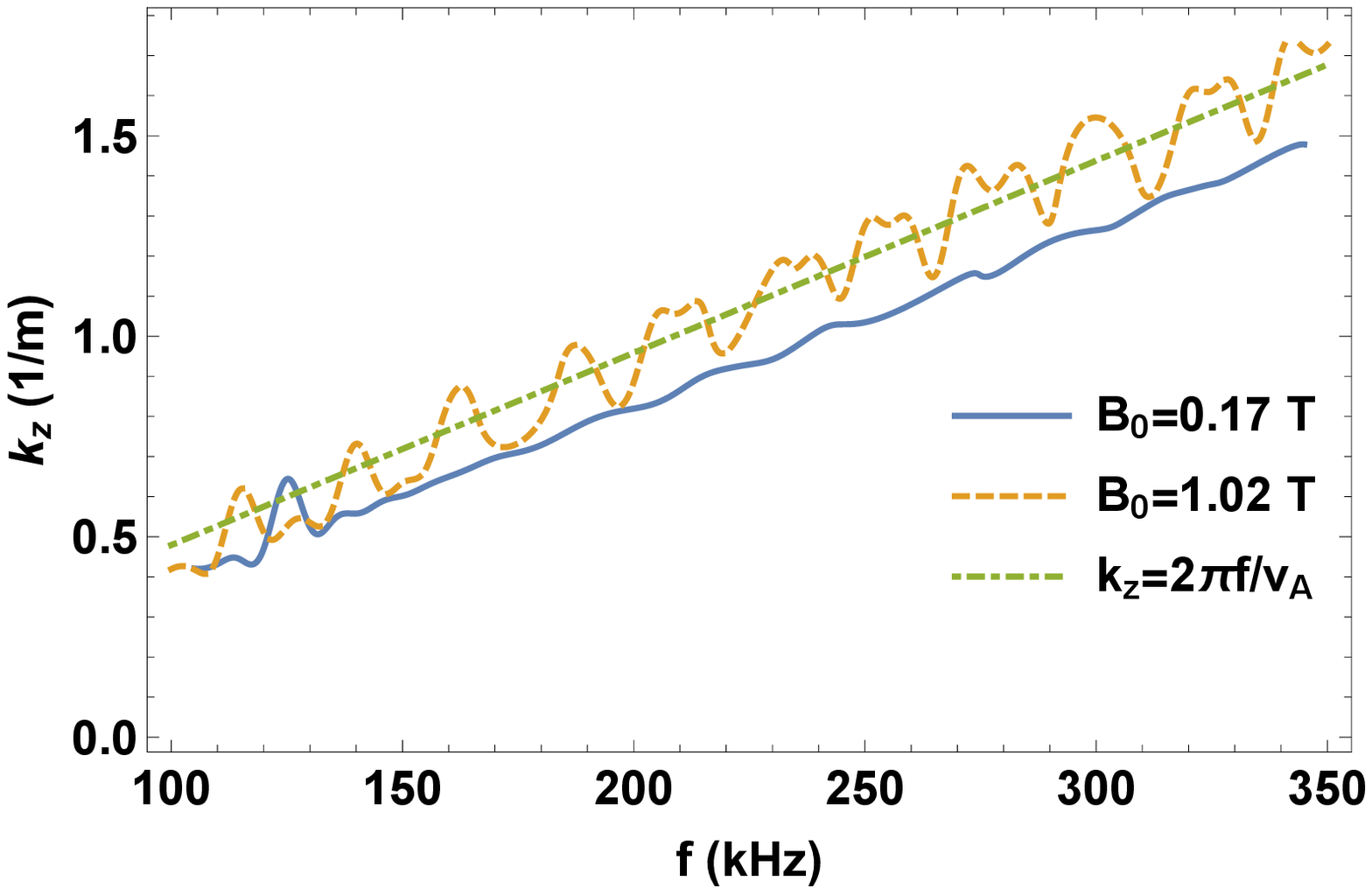}\\
(b)\\
\hspace{-0.1cm}\includegraphics[width=0.46\textwidth,angle=0]{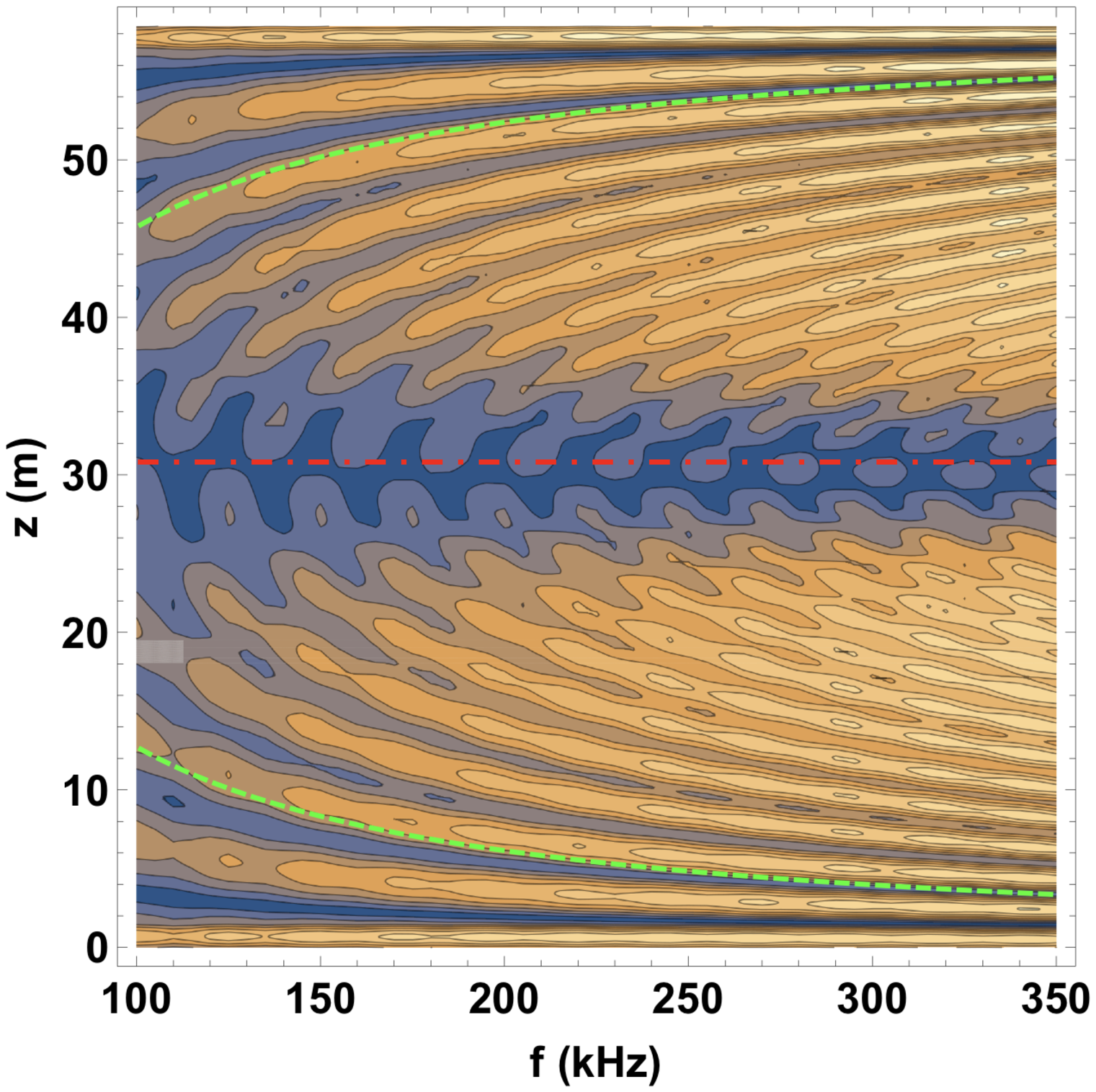}
\end{array}$
\end{center}
\caption{Comparison between numerical wave field and analytical model: (a) dispersion curves of wave field on axis, (b) contour plot of Fig.~\ref{fg4}(d) overlaid with $\lambda=v_A/f$ (green dashed) and the center of antenna (red dot-dashed, $z=30.5$~m).}
\label{fg5}
\end{figure}

Computed results are shown in Fig.~\ref{fg3} and Fig.~\ref{fg4} for $B_0=0.17$~T and $B_0=1.02$~T respectively, including the typical wave field structure in radial and axial directions for both single and multiple frequencies. The radial profiles of real and imaginary components of $B_\theta$ in Fig.~\ref{fg3}(a) and Fig.~\ref{fg4}(a) show a short decay length, as expected from the steep radial density profile employed. The radial variations of $B_{\theta rms}$ magnitude at different axial locations shown in Fig.~\ref{fg3}(b) and Fig.~\ref{fg4}(b) illustrate the effect of magnetic throat on wave field structure: radial wave structure narrows and peaks when approaching the strong magnetic throat. These smooth radial profiles of wave field also imply the absence of continuum damping resonance for Alfv\'{e}nic waves, which is beneficial for gap eigenmode formation. 
Comparing Fig.~\ref{fg3}(c) with Fig.~\ref{fg4}(c), one could see that the axial variation of $B_\theta$ for $B_0=0.17$~T shows a shorter decay length than that for $B_0=1.02$~T, and as a result has a much stronger magnitude near the driving antenna than the latter. Shorter decay length leads to more wave energy deposited near the antenna, and bigger depth of magnetic throat results in relatively stronger wave energy accumulation near the ending throat, as shown in Fig.~\ref{fg3}(d) and Fig.~\ref{fg4}(d), which is another interesting effect of magnetic throat on Alfv\'{e}nic waves.
The dispersion relations of solved wave field are given in Fig.~\ref{fg5}(a), where the simple relation of $k_z=2\pi f/v_A$ for shear Alfv\'{e}nic mode in a slab geometry is also overlaid. The numerical data are obtained by running the EMS for various driving frequencies, selecting the $m=0$ azimuthal component of the plasma response, and calculating the dominant axial wavenumber $k_z$ of this component via Fourier decomposition. The appeared wiggles in the dispersion curve for $B_0=1.02$~T are caused by reflections from endplates, which can be smoothed when the plasma cylinder is made much longer (for example $90$~m). The wave field for $B_0=0.17$~T has much shorter decay length thus weaker reflections, and the dispersion curve is less wiggled. However, overall, higher field strength gives better agreement with the analytical curve due to much longer decay length. This agreement is also confirmed by the contour plot of Fig.~\ref{fg4}(d), as illustrated in Fig.~\ref{fg5}(b). The distance between wave front and the near endplate, which stands for the wave length, decreases when the frequency is increased and exhibits a hyperbolic shape. Recalling the analytical dispersion relation of $\lambda=v_A/f$ (equivalent to $k_z=2\pi f/v_A$), we plot this relation directly using on-axis conditions and overlay the obtained analytical curves with the contour plot and find that they agree quite well. Please note that the lower branch of green line is a direct visualization of $\lambda(f)$ function in an equivalent form of $z(f)$, because $\lambda$ is proportional to $z$ coordinate labelling the distance between wavefront and the left end of machine ($z=0$~m), and the upper branch of green line is a mirrored result regarding the location of antenna (red line).

\subsection{Periodic equilibrium field with local defect}
\begin{figure}[ht]
\begin{center}
\includegraphics[width=0.45\textwidth,angle=0]{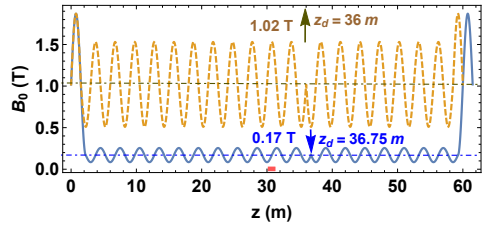}
\end{center}
\caption{Two configurations of periodic equilibrium field with magnetic throats and local defects: $B_0(z)=0.17\times[1+0.5\cos(2\pi z/3)]$~T and $B_0(z)=1.02\times[1+0.5\cos(2\pi z/3)]$~T with the defect located at $z_d=36.75$~m and $z_d=36$~m, respectively. The red bar labels the position of antenna. }
\label{fg6}
\end{figure}

\begin{figure*}[ht]
\begin{center}$
\begin{array}{ll}
(a_1)&(b_1)\\
\includegraphics[width=0.45\textwidth,angle=0]{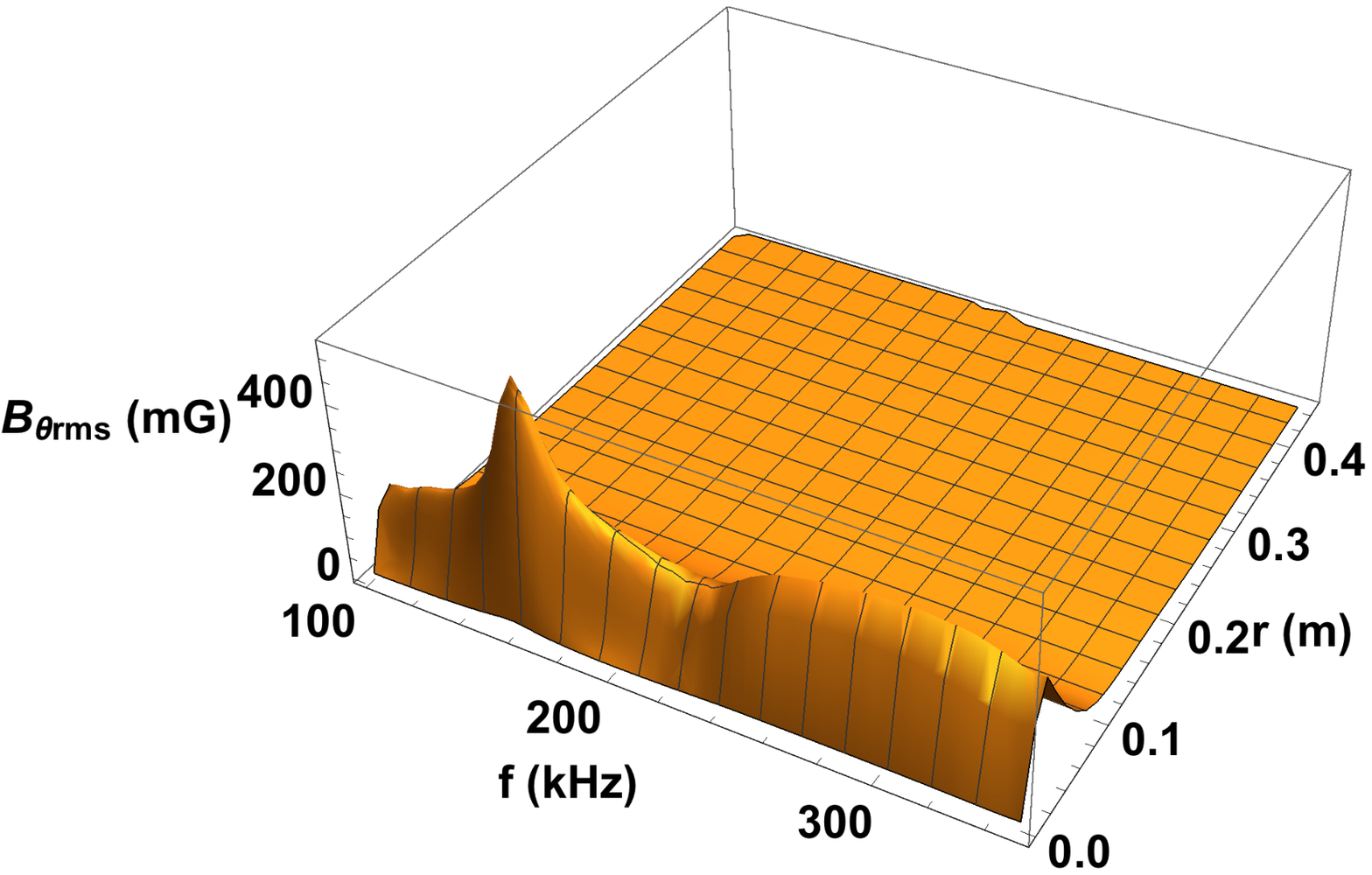}&\includegraphics[width=0.45\textwidth,angle=0]{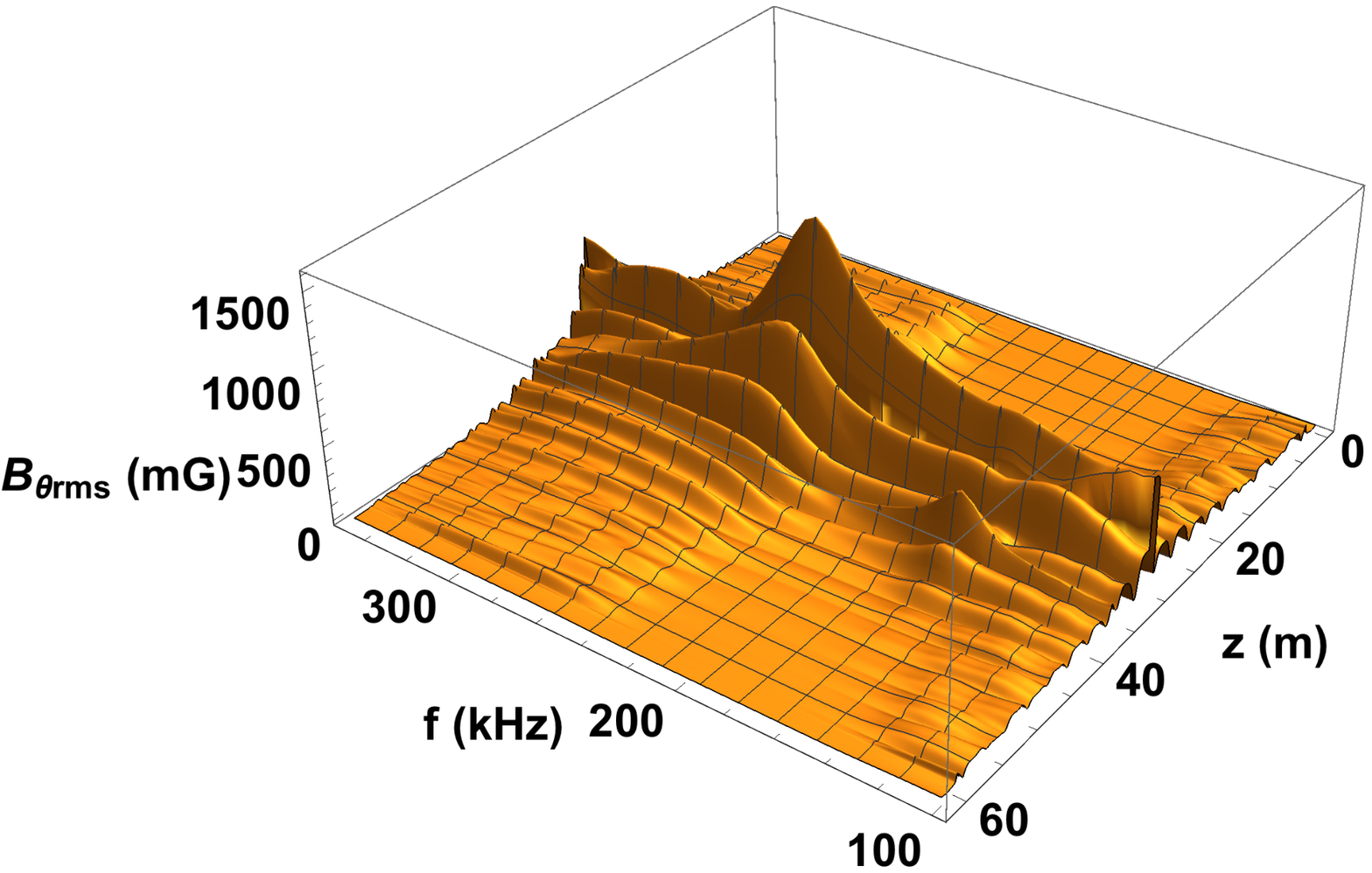}\\
(a_2)&(b_2)\\
\includegraphics[width=0.45\textwidth,angle=0]{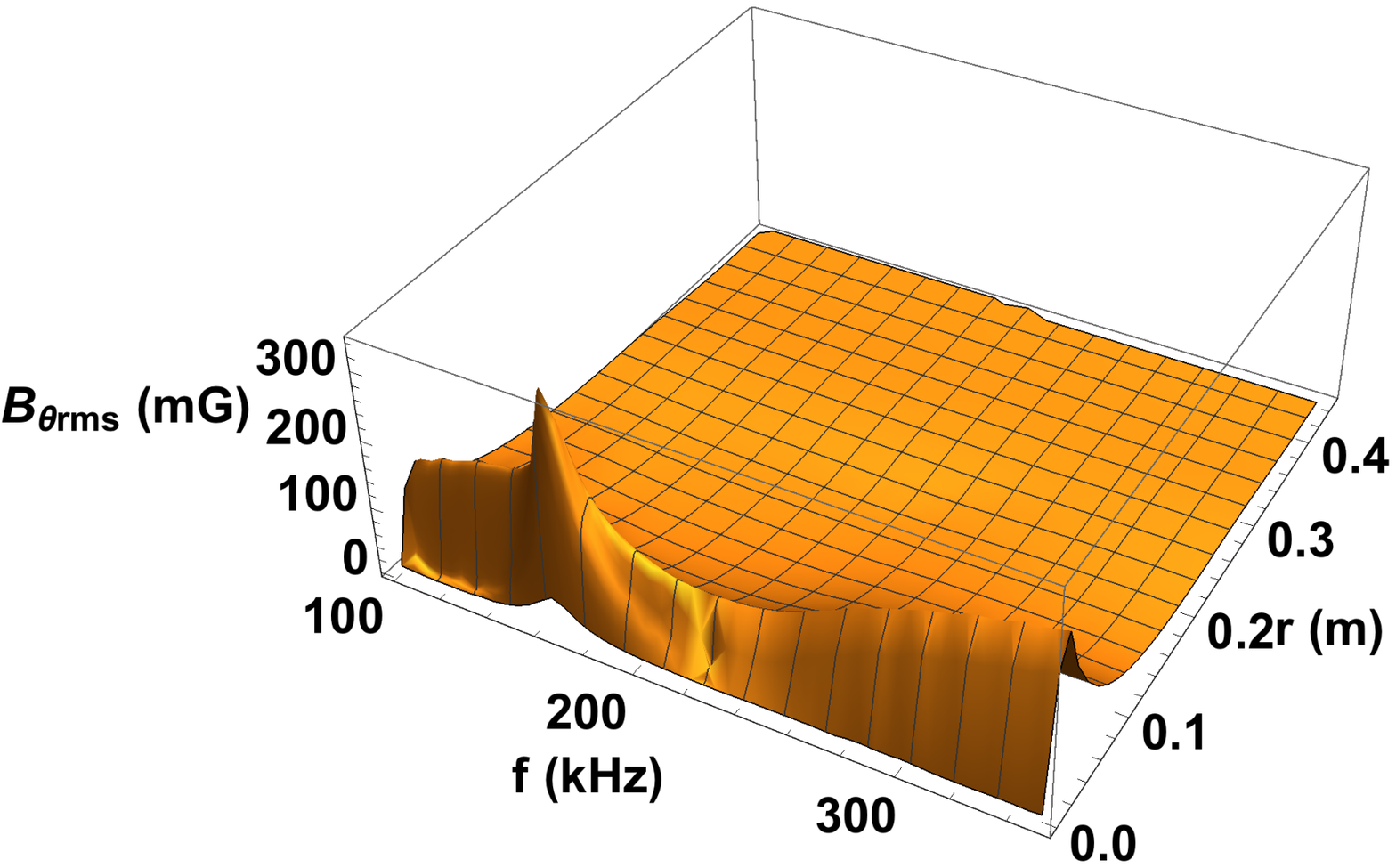}&\includegraphics[width=0.45\textwidth,angle=0]{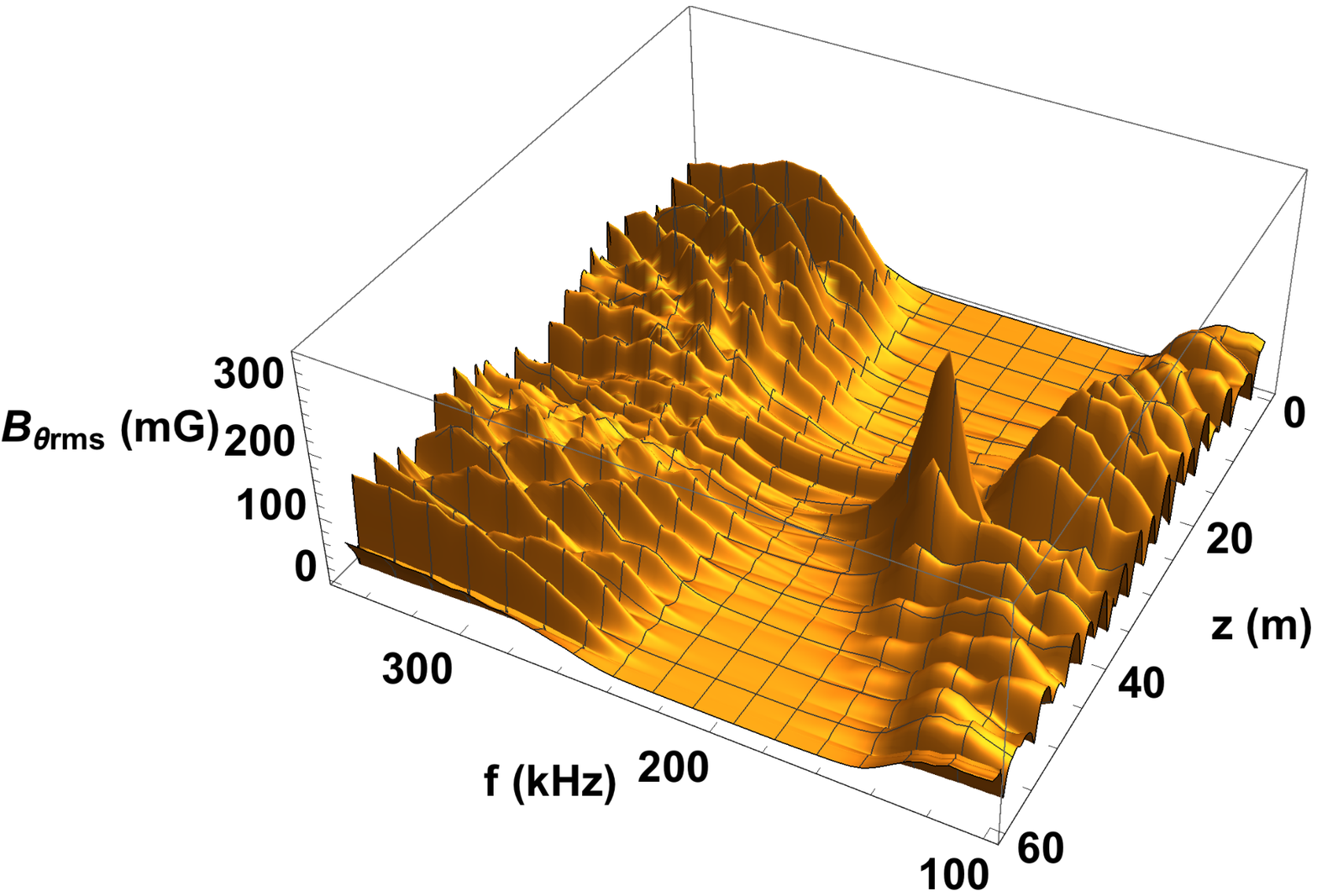}
\end{array}$
\end{center}
\caption{Surface plots of wave magnetic field in the spaces of $(f;~r)$ and $(f;~z)$: ($a_1$) $B_0(z)=0.17\times[1+0.5\cos(2\pi z/3)]$~T, $z=36.75$~m; ($b_1$) $B_0(z)=0.17\times[1+0.5\cos(2\pi z/3)]$~T, $r=0.0195$~m; ($a_2$) $B_0(z)=1.02\times[1+0.5\cos(2\pi z/3)]$~T, $z=36$~m; ($b_2$) $B_0(z)=1.02\times[1+0.5\cos(2\pi z/3)]$~T, $r=0.0075$~m. }
\label{fg7}
\end{figure*}

Following the similar strategy as utilized in previous studies,\cite{Chang:2013aa, Chang:2014aa, Chang:2016aa, Chang:2018aa} a local defect is introduced to the otherwise perfect periodic system, as shown in Fig.~\ref{fg6}. The magnetic ripples between ending throats are constructed by $B_0(z)=0.17\times[1+0.5\cos(2\pi z/3)]$~T and $B_0(z)=1.02\times[1+0.5\cos(2\pi z/3)]$~T with the defect located at $z_d=36.75$~m and $z_d=36$~m, respectively. This defect configuration corresponds to the boundary condition of $E_\theta(z_d)=0$ and produces odd-parity gap eigenmode, phrasing the parity of real and imaginary components of $E_\theta(z)$ across the defect location.\cite{Chang:2013aa, Chang:2014aa} The other defect configuration for $E_\theta'(z_d)=0$ was also conducted and even-parity gap eigenmode was obtained with similar features as shown in Fig.~\ref{fg7}. 
The formed AGEs are shown in Fig.~\ref{fg7} for both $(f;~r)$ and $(f;~z)$ spaces. Here the effective collisionality (sum of original Coulomb collisions and electron Landau damping) has been decreased by a factor of $10$ for clear resonance peak observation. The eigenmode strength drops and width broadens as the collisionality is increased, but is still clearly visible even at the highest collision frequency, as observed before.\cite{Chang:2013aa} The strongest AGEs are $527$~mG ($f=160$~kHz, $r=0.0195$~m, $z=36.75$~m) for $B_0(z)=0.17\times[1+0.5\cos(2\pi z/3)]$~T and $343$~mG ($f=165$~kHz, $r=0.0075$~m, $z=36$~m) for $B_0(z)=1.02\times[1+0.5\cos(2\pi z/3)]$~T. These are in order of $3.1\times10^{-4}$ and $3.4\times10^{-5}$ to the equilibrium field, respectively, making them easily observable in experiment.\cite{Zhang:2008aa} Relatively the resonance peak in Fig.~\ref{fg7}($b_2$) is much more clear than that in Fig.~\ref{fg7}($b_1$), which can be attributed to the higher field strength that yields longer decay length of Alfv\'{e}nic waves in Fig.~\ref{fg4}($d$) than in Fig.~\ref{fg3}($d$). 

\section{Gap eigenmode dependence}\label{dpd}
\subsection{Number of magnetic ripple}
\begin{figure}[ht]
\begin{center}$
\begin{array}{l}
(a)\\
\includegraphics[width=0.45\textwidth,angle=0]{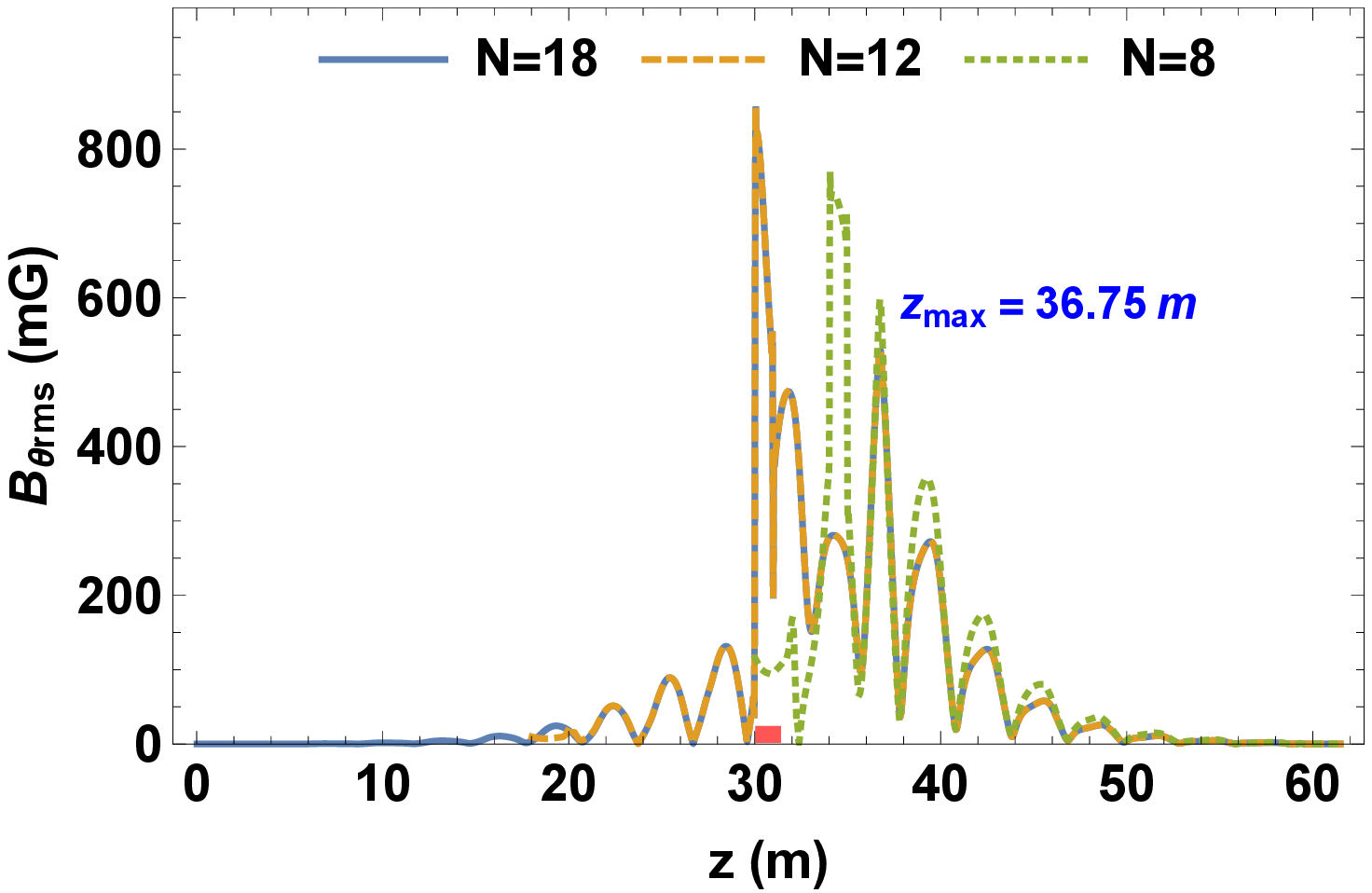}\\
(b)\\
\includegraphics[width=0.45\textwidth,angle=0]{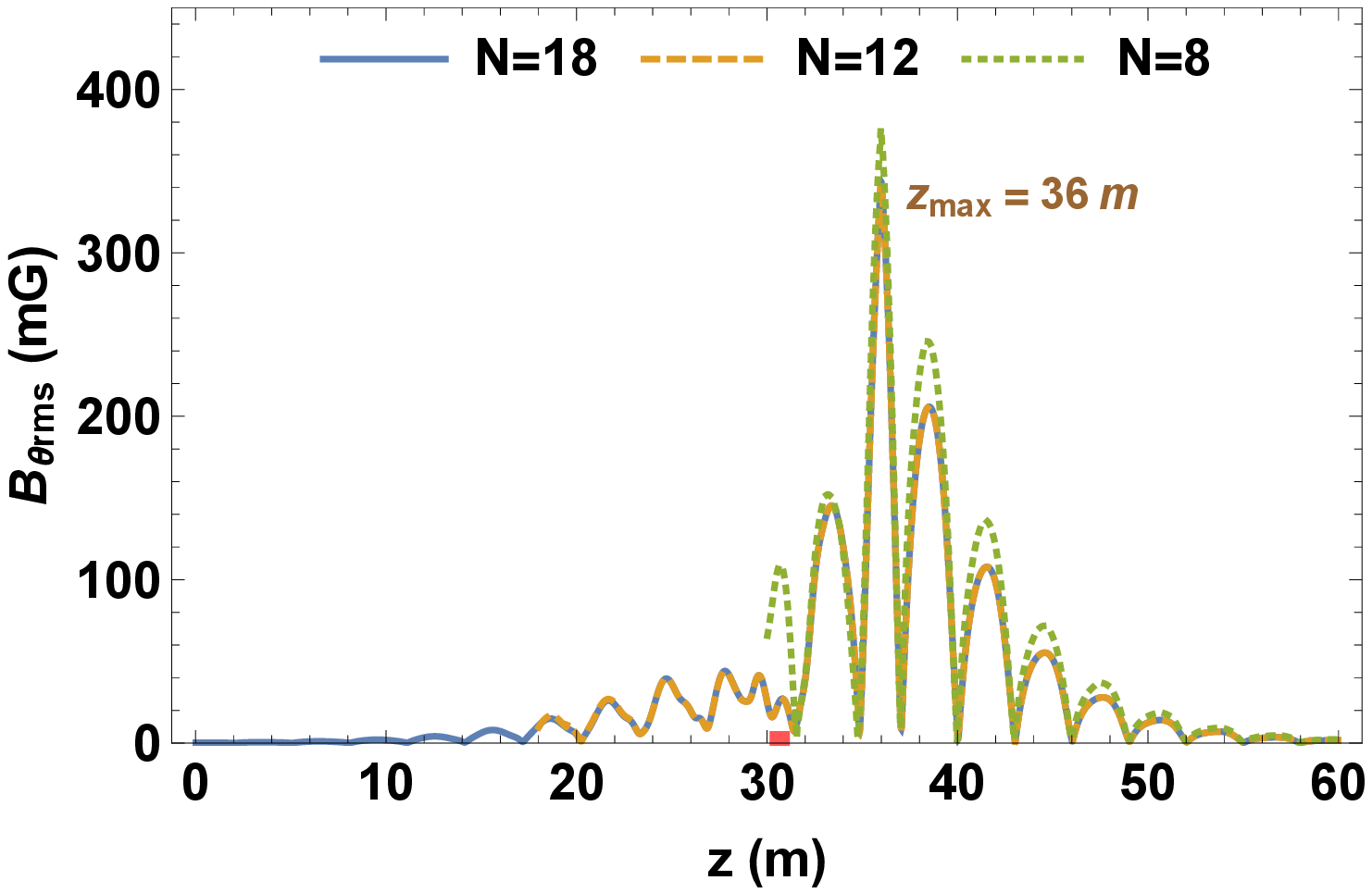}
\end{array}$
\end{center}
\caption{Axial variations of strongest AGE for different numbers of magnetic ripples: (a) for $B_0(z)=0.17\times[1+0.5\cos(2\pi z/3)]$~T at $f=160$~kHz and $r=0.0195$~m, (b) for $B_0(z)=1.02\times[1+0.5\cos(2\pi z/3)]$~T at $f=165$~kHz and $r=0.0075$~m. The red bar labels the position of antenna.}
\label{fg8}
\end{figure}

Because the machine length around $60$~m and ripple depth of $0.5$ are costly and challenging to implement in experiment, it is highly desirable to investigate the gap eigenmode dependence on the number ($N$) and depth ($\epsilon$) of magnetic ripples. First, we gradually cut off the ripples left to the antenna, and run EMS for shortened plasma cylinder with all other conditions unchanged. Figure.~\ref{fg8} shows the axial variations of strongest AGE for three typical values of $N$: $18$, $12$, $8$. It can be seen that the gap eigenmode remains nearly the same when $N$ is reduced from $18$ (Fig.~\ref{fg6}) to $12$ but becomes a little stronger for $N=8$. However, the wavelength of AGE is always twice the system's period, which is consistent with Bragg's law,\cite{Kittel:1996aa} and the magnitude always peaks at the defect location.

\subsection{Depth of magnetic ripple}
\begin{figure}[ht]
\begin{center}$
\begin{array}{l}
(a)\\
\includegraphics[width=0.45\textwidth,angle=0]{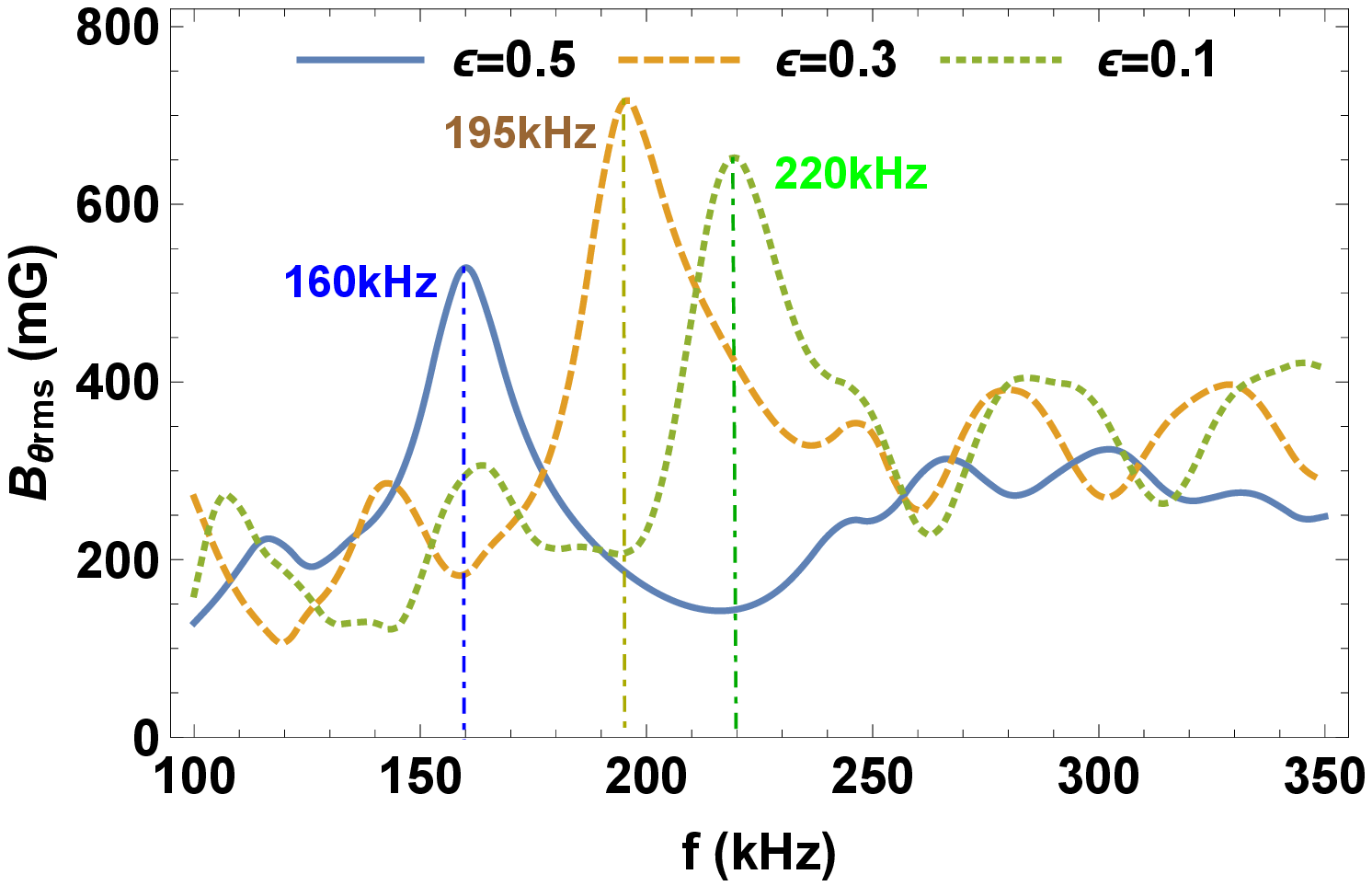}\\
(b)\\
\hspace{-0.15cm}\includegraphics[width=0.458\textwidth,angle=0]{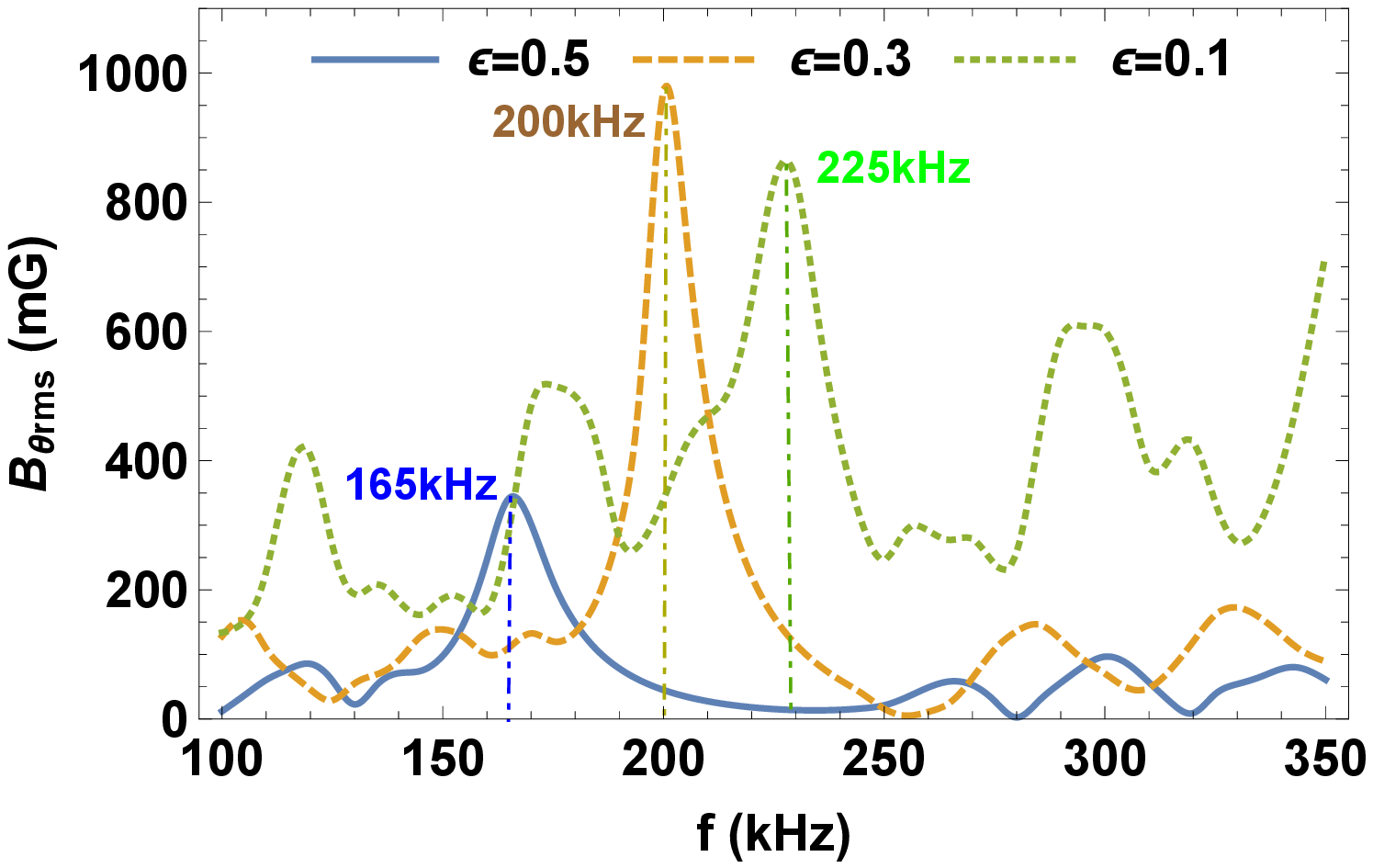}
\end{array}$
\end{center}
\caption{Dependence of strongest AGE on the depth of magnetic mirrors: (a) for $B_0(z)=0.17\times[1+\epsilon\cos(2\pi z/3)]$~T at $z=36.75$~m: $160$~kHz at $r=0.0195$~m (blue), $195$~kHz at $r=0.0225$~m (brown), $220$~kHz at $r=0.0275$~m (green); (b) for $B_0(z)=1.02\times[1+\epsilon\cos(2\pi z/3)]$~T at $z=36$~m: $165$~kHz at $r=0.0075$~m (blue), $200$~kHz at $r=0.0135$~m (brown), $225$~kHz at $r=0.0225$~m (green).}
\label{fg9}
\end{figure}

Then we keep $N=12$ and vary $\epsilon$ from $0.5$ to $0.1$. Figure.~\ref{fg9} shows the resulted dependence of strongest AGE for three typical values of $\epsilon$: $0.5$, $0.3$, $0.1$. It seems that there exists a certain value of depth, for which the eigenmode magnitude is maximum, for example around $\epsilon=0.3$. More importantly, an upward frequency shift can be seen clearly as the depth of magnetic ripple drops, agreeing with a previous observation.\cite{Chang:2016aa} Although detailed physical explanation has not yet been given, it is consistent with a recent finding that the whole spectral gap moves to higher frequency region as the depth is reduced.\cite{Chang:2018ab} Recalling that this modulation depth is proportional to the width of spectral gap ($\bigtriangledown f/f_0=\epsilon$),\cite{Zhang:2008aa, Chang:2013aa, Chang:2014aa} this seems be a flute-like effect that the resonant cavity (spectral gap) shrinks so that the gap eigenmode frequency grows, similar to the tone tuning of a flute by shortening the resonant air column. 

Therefore, it suggests that a linear plasma machine with $8$ magnetic ripples (or length of $30$~m) and modulation depth of $0.3$ is enough and best for the AGE observation. 

\section{Discussion and summary}\label{smr}
For experimental implementation of this AGE, there are four possible issues to concern: (i) the dissipative processes in the plasma will not destroy the gap-mode resonance (e. g. through varying electron temperature and plasma radius), (ii) there should be an axial slot in the conducting ring if employed as a local defect to prevent the formation of induced azimuthal current, (iii) the radial density profile if controllable does not yield continuum damping resonance of Alfv\'{e}nic mode, and (iv) the range of frequency scan need be broad enough to cover the possible frequency shift when varying the depth of magnetic ripples. However, there is no visible fundamental barrier that is unsolvable in experiment, and given that the strength of formed AGE can be in order of $3.1\times10^{-4}$ to the equilibrium field strength, this observation is very promising referring to the diagnostic results in \cite{Zhang:2008aa}. The effect of this gap eigenmode on particle transport is that once driven to large amplitude the mode can change the particle orbit of motion, through resonances between wave phase velocity and orbital frequencies, and eventually change the transport rate.\cite{Berk:1996aa, Heeter:2000aa, Collins:2016aa, Qiu:2018aa} Unfortunately, quantitive calculations using measured mode amplitudes underestimate the observed fast-ion transport at present, and detailed identification of nonlinear transport mechanism is immature in experiment.\cite{Heidbrink:2008aa} This paper is critical for the design of a linear low-temperature plasma device with remarkable diagnostic resolutions in both space and time, which aims to study the interaction physics between AGE and energetic particles and is believed to be capable of improving our understanding of the involved particle transport mechanism significantly.

In summary, the present work computes Alfv\'{e}nic waves propagating in a linear plasma ended with magnetic throats, which are designed to confine the axial movement of charged particles. Once confined the energetic ions can interact with AGE many times as happened in toroidal fusion plasmas. The radial and axial variations of wave field for uniform equilibrium field between magnetic throats are presented, for both single and multiple driving frequencies, showing a robust structure which is free of continuum damping resonance and agrees well with shear Alfv\'{e}nic branch. The effects of magnetic throat on Alfv\'{e}nic waves include narrowing radial mode structure, enhancing mode strength, and accumulating wave energy. For periodic equilibrium field with local defect, AGE is clearly formed inside spectral gap. It is a standing wave localized around the defect location and has wavelength twice the system's period, consistent with Bragg's law. This mode behaves insensitive to the reduction of magnetic ripples from $18$ to $8$, but experiences an upward frequency shift when the depth of magnetic mirrors drops from $0.5$ to $0.1$. The latter may be caused by a flute-like effect that the shrinking resonant cavity (namely spectral gap) forces the mode frequency to increase, similar to the upward tone tuning of a flute when the air column is shortened. This work is of great practical interest for designing a linear plasma machine to form AGE and further study its interaction with confined energetic ions, together with the nonlinear behavior and turbulence of Alfv\'{e}nic waves and other relevant frontiers of physics.

\acknowledgements
Inspiring and helpful discussions with Ran Chen and Ming Xu are appreciated. This work is supported by various funding sources: National Natural Science Foundation of China (11405271), China Postdoctoral Science Foundation (2017M612901), Chongqing Science and Technology Commission (cstc2017jcyjAX0047), Chongqing Postdoctoral Special Foundation (Xm2017109), Fundamental Research Funds for Central Universities (YJ201796), Pre-research of Key Laboratory Fund for Equipment (61422070306), and Laboratory of Advanced Space Propulsion (LabASP-2017-10). 

\section*{References}
\bibliographystyle{unsrt}

\end{document}